\def\Napoli{Dipartimento di Scienze Fisiche, Universit\`{a} "Federico II" di
Napoli "Federico II" and Istituto Nazionale di Fisica Nucleare,
sezione di Napoli , Complesso Universitario di Monte S. Angelo,
via Cintia, 80126 Napoli, Italy}
\def\Lebedev{P.N. Lebedev Physical Institute, Moscow}
\def\Murcia{Departamento de Matem\'{a}tica Aplicada, Facultad de Inform\'tica \\
Campus de Espinardo, 30100 Murcia, Spain}
\def\nn{\nonumber}

\def\w{\omega}
\def\ni{\noindent}
\def\calG{{\cal G}}
\def\calP{{\cal P}}
\def\calO{{\cal O}}
\def\calF{{\cal F}}

\def\calD{{\cal D}}
\def\calS{{\cal S}}
\def\calL{{\cal L}}
\def\calH{{\cal H}}

\def\medio{\frac{1}{2}}

\def\parcial#1{\frac{\partial}{\partial #1}}

\documentstyle[12pt]{article}

\begin{document}

\begin{center}
{\Large {\bf Geometrical Aspects of Lie Groups Representations
and Their Optical Applications}}
\end{center}

\bigskip

\centerline{ J. Guerrero$^{1,2}$, V. I. Man'ko$^{3}$, G. Marmo$^{2}$ and
A. Simoni$^2$}

\bigskip
\bigskip

\footnotetext[1]{\Murcia} \footnotetext[2]{\Napoli}
\footnotetext[3]{\Lebedev}

\begin{center}
{\bf Abstract}
\end{center}

\small

\begin{list}{}{\setlength{\leftmargin}{3pc}\setlength{\rightmargin}{3pc}}
\item In this paper we present a new procedure to obtain unitary and
irreducible
representations of Lie groups starting from the cotangent bundle of the
group (the cotangent group).
We discuss some applications of the construction in quantum-optics problems.
\end{list}

\normalsize

\vskip 1cm

\section{Introduction}

Any quantization procedure \cite{Dirac}, i.e., a way to construct
a quantum system starting from a classical one, provides us with
the construction of unitary representations of Lie groups, often
arising in the picture as symmetry groups. Various symmetry groups
(or dynamical symmetry groups \cite{Alberto}) describe the
properties of different quantum systems
\cite{Barut0}\cite{Sud}\cite{Gell}\cite{Malk} and quantum optical
devices. The description of the energy spectrum of a quantum
system like a trapped ion, or the evolution of the two-level atom
in a cavity can be given either in terms of solutions of
stationary and nonstationary Schr\"odinger equations or using
matrix elements of the appropriate irreducible representations of
the dynamical symmetry groups. The Heisenberg--Weyl group ISp(4,R)
and noncompact group SU(1,1) as well as other groups are
well-known examples (see
\cite{Sim}\cite{Klauder1}\cite{Man}\cite{Dod}) which are used to
describe properties of optical and quantum optical phenomena. The
application of  group-theoretical methods in quantum mechanics and
quantum optics is based on using the construction of matrix
elements of irreducible representation of the various Lie groups
and tools like Casimir operators whose eigenvalues determine the
irreducible representations. In view of such an importance of the
Lie group representations in physical applications, it is worth
understanding the aspects of the irreducible representation
constructions as deeply as possible. An elegant approach to the
physical problems as well as to the problems of the dynamical
symmetry groups is the geometrical approach. This approach permits
to give a unified picture of various phenomena in quantum
mechanics and quantum optics expressed in terms of the Lie group
representation theory. The relation of the classical domain to the
quantum one can be associated in this geometrical approach in
terms of the procedure called ``geometrical quantization.'' One of
the goals of this paper is a discussion of some geometrical
aspects of the Lie group representation theory.

The notion of elementary system introduced by Wigner \cite{Wigner} requires
the representation to be irreducible. Therefore, quantization procedures are
often required to provide unitary irreducible representations of Lie
groups. The so called method of orbits, elaborated by Kirillov
\cite{Kirillov} and Kostant \cite{Kostant} on the mathematical side and
by Souriau \cite{Souriau} on the physical side, was an attempt to
geometrize
and extend the quantization procedure used in physics starting with
classical systems defined on phase-space (cotangent bundles of configuration
spaces). The irreducibility problem, tackled by introducing polarizations,
has been one of the major problems in this geometric quantization program.

As, to use the words of A. A. Kirillov \cite{Kirillov}, ``the possibilities
of geometric quantization are still far from having been exhausted", it
seems to us useful to consider what is the connection between the orbit
method, the construction of representations on homogeneous spaces and the
role and implementation of polarizing conditions to select irreducible
representations.

Because the orbit method uses orbits of the co-adjoint action of a
Lie group, say $G$, on the dual of its Lie algebra, say $\calG^*$,
and homogenous spaces are usually realized as quotients $G/K$,
with $K$ a closed subgroup of $G$, it is useful to consider these
spaces as arising from $T^*G$ which is the ``phase-space" on the
configuration space $G$, using therefore a notion closer to the
physicists view of quantization for classical systems.

Our paper is organized in the following way. We first consider geometrical
entities on the cotangent bundle $T^*G$ of any Lie group $G$. Here we
show the important role that Poisson subalgebras play in going from
$T^*G$ to $\calG^*$. More generally we show that Poisson subalgebras
are connected with quotient spaces that make the quotienting  map a Poisson
map. The left and right action of $G$ on itself, when extended to $T^*G$,
provide a link with the problem of irreducibility of representations.
The possibility of constructing the ``quantization map" in terms of
canonical variables, with one eye on the irreducibility problem, is tackled
via a generalized Jordan-Schwinger map.

The geometrical picture emerging from these considerations will show the
limitations of the construction with respect to the problem of locality
versus globality, which is connected with the problem of self-adjointness
of the constructed operators \cite{Klauder}. Here we should stress that
usually problems of (common) domains of the operators constructed by
quantization are not addressed.

The irreducible representations of classical Lie groups were
constructed (see, e.g. \cite{Mackey}) in a purely mathematical
context. On the other hand in physical applications it is
desirable to consider the construction of irreducible
representations as close as possible to the experience of
theoretical physics. One of the aspects of this experience is the
use of classical Mechanics methods, with the notion of phase
space, Poisson brackets, and the notion of canonical quantization
of the classical systems for which the classical positions and
momenta become the operators acting in a Hilbert space of states
of the quantum system. An attempt to find the link of the
``physical picture" of the irreducible representations
construction with known mathematical formulae for the irreducible
representations of Lie groups was done in \cite{LMM}. The
considerations in this work was based on the properties of the
regular representation of the Lie group treated as classical
motion in the ``phase space" of the group, but this consideration
did not use the geometrical point of view. Here we shall try to
elucidate a geometrical picture of the group phase space
construction and the quantization picture for the known
irreducible representations of Lie groups.

Important known example of the link of the group representation construction
with the physical system which is the two-mode quantum oscillator is the
Jordan-Schwinger map \cite{Jordan,Schwinger}. This map was used to express the
generators of the irreducible representations of the compact group $SU(2)$
in terms of the creation and annihilation operators of the two-mode
oscillator.

Geometrical properties of the Jordan-Schwinger map and its
extension to quantum groups were discussed in \cite{MZMV}. Using
the experience of the Jordan-Schwinger map construction we would
like to generalize and apply it for considering generic Lie groups
representations. Thus the aim of this paper is to give a
geometrical interpretation for the Lie group ``phase space"
discussed in \cite{LMM} and the irreducible representations. We
want to treat the group phase space as the known geometrical
object called the cotangent bundle $T^*G$. Also we suggest a
conjecture that all the irreducible representations of the group
$G$ can be associated with the orbits of the group in the group
phase space, representing the cotangent bundle $T^*G$ as the union
of the orbits obtained due to right and left action of the group
on the group phase space. It is important that the generators of
the right and left regular representation can be considered for
generic orbits as linear in momenta functions. Of course, the
geometrical construction of group representations is a very well
known procedure. But usually the construction is based on
considering a fixed orbit \cite{Kostant,Souriau}, or on
considering the group \cite{Aldaya}. The novelty of the present
geometrical interpretation of the irreducible representations is
in studying the extended object $T^*G$ as the phase space of a
dynamical system analog containing information on the irreducible
representations of the Lie group. It should be noted that the
construction of \cite{LMM} was used \cite{Leznov} to consider new
types of integrable systems.

\section{Lie groups and their cotangent bundles}

 In this section we construct bundles over a Lie group $G$ which are themselves
Lie groups.
For simplicity of notation we are assuming our Lie groups to be realized in
terms of matrices, however this assumption does not play any role in our
considerations.

 Given any Lie group we define on it left and right invariant 1-forms by
introducing a basis for the Lie algebra ${\cal G}$ of $G$, say
$(\tau^1,\tau^2,\ldots,\tau^n)$, $n={\rm dim} G$, and setting:
\begin{equation}
g^{-1}dg = \theta^L_j\tau^j\,,  \qquad\qquad dg\ g^{-1} = \theta^R_j\tau^j\,,
\end{equation}

\ni where $g\in G$ and $\theta^L$ and $\theta^R$ are
the left invariant and right invariant 1-forms, respectively.


Because $g^{-1}dg= g^{-1}(dg\ g^{-1})g$, we find
$\theta^L_j\tau^j=g^{-1}\tau^k g\ \theta^R_k$, and by writing the
adjoint action as
\begin{equation}
Ad(g)_j^k\tau^j = g^{-1}\tau^k g\,,
\end{equation}

\ni we find $\theta^L_j=Ad(g)_j^k\ \theta^R_k$, i.e. left and right invariant
1-forms are connected by the adjoint transformation. We also have
$Ad(g^{-1})^k_j\ \theta^L_k=\theta^R_j$.

By using these 1-forms we construct the left invariant and right invariant
volume elements, $\Omega^L=\theta^L_1\wedge\theta^L_2\wedge \ldots \wedge
\theta^L_n$ and $\Omega^R=\theta^R_1\wedge\theta^R_2\wedge \ldots \wedge
\theta^R_n$, respectively. When $\Omega^L=\Omega^R$ the Lie group $G$ is
said to be unimodular.

 By using the Riesz representation theorem we can associate a measure
with either one of the two volumes. In any case we set
\begin{equation}
<\Psi|\Phi>^L = \int_G \Psi^*\Phi\ \Omega^L\,,
\end{equation}

\ni and
\begin{equation}
<\Psi|\Phi>^R = \int_G \Psi^*\Phi\ \Omega^R \,.
\end{equation}

The left action of $G$ on itself preserves $\Omega^L$ and defines the
left regular representation of $G$, and similarly for the right representation.

 By using infinitesimal generators of the right and left action, $X^L$ and
$X^R$, respectively, we find:
\begin{equation}
\theta^L_j((X^L)^k)=\delta_j^k\,,\qquad\qquad \theta^R_j((X^R)^k)=\delta_j^k\,,
\end{equation}

\ni i.e. right generators $X^L$ are left-invariant and left generators
$X^R$ are right-invariant. Clearly $\theta^L_j((X^R)^k)=Ad(g)_j^l\delta_l^k=
Ad(g)_j^k$.

 For any group $G$, we can construct the tangent bundle $TG$ and the
cotangent bundle $T^*G$, these two bundles carry a Lie group
structure associated with the adjoint representation and the
co-adjoint representation respectively. On the one hand, if the
group $G$ carries a natural metric, it is possible to define a non
degenerate Lagrangian function on $TG$ and consider associated
geodetical motions along with symmetries and constants of the
motion. It is possible to define a Lagrangian symplectic structure
and Lagrangian Poisson brackets, i.e. these structures are tied to
a specific Lagrangian dynamics. On $T^*G$, on the other hand, it
is always possible to define a symplectic structure along with
canonical actions of $G$, left and right actions, with
``Hamiltonian generating functions" closing on the Lie algebra of
$G$ in terms of Poisson brackets \cite{MSSV}.

The existence of a left invariant or right invariant volume element on $G$
allows to define a measure on $G$, the Haar measure, which can be used to
define a Hilbert space of square integrable functions on $G$. On this Hilbert
space $G$ has a unitary representation along with a decomposition into
irreducible components. It is possible to go from one Hilbert space
to the other by using the correspondence between left action and right action.
The canonical action of $G$ on $T^*G$, allows to consider these Hilbert
spaces realized in terms of functions on $T^*G$ bringing in the possibility
of using symplectic and Poisson geometry in connection with the decomposition
of unitary representation into irreducible ones.

This interplay is the hard core of Geometric Quantization and any other
proposed quantization where geometry  plays a relevant role.

In addition to previous considerations we also notice that $T^*G$, having
``canonical" coordinates, may be thought of as a non-abelian generalization
of $T^*R^n$ along with its canonical structure of Heisenberg group,
therefore we may try to generalize the Jordan-Schwinger map to the
cotangent bundle of any Lie group.

By using left generators or right generators we can decompose $TG$ into
$G{\times}_R {\cal G}$ and $G{\times}_L {\cal G}$, respectively, and similarly
decompose $T^*G$ into $G{\times}_L {\cal G}^*$ or $G{\times}_R {\cal G}^*$
by using 1-forms.

Given any function $f:T^*G \rightarrow \calG^*$ which is a submersion onto,
we can define two 1-forms on $T^*G$, $\theta^L_f = f(g^{-1}dg)$ and
$\theta^R_f=f(dg\ g^{-1})$. These 1-forms are the potential for
symplectic (non-degenerate) 2-forms $d\theta^L_f$ and $d\theta^R_f$. When $f$
is the identity map with respect to the splitting
$T^*G\approx G{\times}_L \calG^*$ or $T^*G\approx G{\times}_R \calG^*$, we find
the canonical 1-form $\Theta_0$ and we have
\begin{equation}
\Theta_0= \calP^L(g^{-1}dg)=\calP^R(dg\ g^{-1})\,,
\end{equation}

\ni where $\calP^L,\calP^R:T^*G\rightarrow \calG^*$ are the left (right)
invariant momenta which are the generating functions of the canonical
right (left) actions of $G$ on $T^*G$.

 Let us denote by $\tilde{X}^L$ ($\tilde{X}^R$) the infinitesimal canonical
generators of the right (left) action of $G$ on $T^*G$, having generating
functions the momenta $\calP^L$ ($\calP^R$).

 From $(\calP^L)^j\theta^L_j = (\calP^R)^j\theta^R_j = \Theta_0$, we find
$(\calP^L)^jAd(g)_j^k\theta^R_k = (\calP^R)^k\theta^R_k$, i. e.
$(\calP^L)^jAd(g)_j^k = (\calP^R)^k$, and also $(\tilde{X}^L)^j Ad(g)_j^k =
(\tilde{X}^R)^k$.

 The symplectic volume, $\Omega_0=(d\Theta_0)^n = d\Theta_0\wedge \ldots
\wedge d\Theta_0$, can be decomposed into
\begin{equation}
d(\calP^L)^1\wedge \ldots \wedge d(\calP^L)^n\wedge \Omega^L\,,
\end{equation}

\ni or in
\begin{equation}
d(\calP^R)^1\wedge \ldots \wedge d(\calP^R)^n\wedge \Omega^R\,,
\end{equation}

\ni respectively.

As for Poisson brackets, which in matrix notation can be written as
$g^{-1}\{\calP^L,g\} = I$, if we use Gaussian
coordinates for the Lie group $G$, say $(\xi_1,\xi_2,\ldots,\xi_n)$ are the
parameters of the 1-parameter subgroups associated with
$((X^L)^1,(X^L)^2,\ldots,(X^L)^n)$, we find
$\{(\calP^L)^j,\xi_k\} = \delta^j_k$. Of course Gaussian coordinates only
exist in a neighborhood of the identity (or, by translations, in the
neighborhood of a point), therefore they do not capture the global aspects
of our treatment.

 By using the action (left and right) $G{\times} T^*G \rightarrow T^*G$, we
find orbits of $G$ diffeomorphic to $G$ and giving a quotient space
$T^*G/G$ diffeomorphic to $\calG^*$. As a matter of fact, if the lifted
action of $G$ to $T^*G$ is defined via $\theta^{L,R}_f$ we find as
projection map $f:T^*G\rightarrow T^*G/G$.

 We stress that no matter which lift from $G$ to $T^*G$ is being used, the
generating functions of the lifted action of $G$ are always linear in the
corresponding momenta in $\calG^*$.

The projection map $f:T^*G\rightarrow T^*G/G \equiv \calG^*$ is a Poisson
map with respect to the natural Poisson bracket available on $\calG^*$
by using the identification $\calG^*\equiv Lin(\calG,R)$ or
$\calG=(\calG^*)^*$.

Casimir functions on $T^*G$ are simply the inverse image of
functions on $\calG^*$ which are central elements with respect to
the natural Poisson bracket on $\calG^*$. They can also be defined
as those functions on $T^*G$ which commute with $\calP^L$ or
$\calP^R$ (then, as a result of previous analysis, they commute
with both). We also have $\{\calP^L,\calP^R\}=0$, along with
$[X^L,X^R]=0$ and $[\tilde{X}^L,\tilde{X}^R]=0$.

\section{Digression: Manifolds and algebras of functions}

As it is well known, manifolds can also be described by
commutative rings of functions defined on them, i.e. $M=Hom_{\cal
A} (\calF,R)$, points of $M$ are identified with algebra
homomorphisms from the algebra $\calF$ to the algebra $R$. When
$M$ is a Lie group $G$, $\calF(G)$ also possesses a Hopf algebra,
capturing the group multiplication property on the manifold $G$.

If there is a smooth projection man, i.e. a submersion onto
$\phi: M\rightarrow N$, the pull-back $\phi^*(\calF(N))$ defines a subalgebra
of $\calF(M)$. Therefore subalgebras of the commutative algebra $\calF(M)$
constitute a generalization of projection maps onto ``quotient manifolds".
When $M$ carries a Poisson structure, we can define, in addition, Poisson
subalgebras. They are subalgebras which are also closed under Poisson brackets.

When $N$ carries in addition a Poisson bracket, say $\{f,h\}_N$ for
$f,g\in\calF(N)$, we say that $\phi$ is a Poisson map with respect to the
Poisson bracket on $M$, if we have:
\begin{equation}
\{\phi^*f,\phi^*g\}_M = \phi^*(\{f,g\}_N)\,,
\end{equation}

\ni for any $f,g\in\calF(N)$. Therefore $\phi^*(\calF(N))$ is a Poisson
subalgebra. Thus Poisson subalgebras are a generalization of Poisson
projection maps. We recall that Poisson subalgebras were considered by Lie
under the name of ``Functionen gruppen" \cite{Lie}.

The reason why subalgebras are generalizations of projection maps
has to do with the fact that not all subalgebras arise from
pull-backs via projection maps. To identify them we would need
some regularity assumptions on the subalgebras involved.

 In this duality between $M$ and $\calF(M)$, vector fields on $M$ arise
as derivations on $\calF(M)$ and differential forms as $\calF(M)$-multilinear
antisymmetric maps from derivations to $\calF(M)$. Similarly, diffeomorphisms
on $M$ arise as automorphisms of the algebra structure on $\calF(M)$.

Given a subalgebra, say $\calF_N$, in $\calF(M)$, the set of all
derivations of $\calF(M)$ which annihilate $\calF_N$ are a Lie
subalgebra of vector fields, say $\calD^0(N)$. The regularity
assumption on $\calF_N$ can be stated by saying that maximal
integral sub-manifolds of $\calD^0(N)$ define a regular foliation.

\section{The Poisson algebra structure on $\calG^*$}

We recall that on the dual space $\calG^*$ of a Lie algebra $\calG$ we
have a canonical Poisson bracket defined by
\begin{equation}
\{f,h\}(x)=<x|[df(x),dh(x)]>\,,
\end{equation}

\ni where $df\in \calG=(\calG^*)^*$ and the commutator is taken
in $\calG$, while the natural pairing between $\calG$ and $\calG^*$ is
denoted by $<\ |\ >$.

Casimir functions are defined to be those functions on $\calG^*$ which
Poisson commute with any other function, i.e. the center of the Lie algebra
defined by the Poisson structure on $\calF(\calG^*)$.

The group $G$ acts on $\calG^*$ via the co-adjoint action and
defines the co-adjoint orbits. Each orbit is a symplectic manifold
with symplectic structure given by inverting the previous Poisson
bracket on each orbit. The stability group of a point $x$ in
$\calG^*$ is a closed subgroup of $G$ and will be denoted by
$G_x$, the corresponding Lie algebra will be denoted by $\calG_x$.

For a  generic $x$, $G_x$ is abelian, and generic orbits are
defined as level sets of the Casimir functions. Nongeneric orbits
are level sets of $Ad^*$-invariant relations. They are invariant
sub-manifolds defined by level sets of functions which are
$Ad^*$-invariant only for specific values \cite{Levi-Civita}.

Any element $x\in\calG^*$ defines a left invariant 1-form
$\theta^L_x$ and a right invariant 1-form $\theta^R_x$ on the
group $G$. With these 1-forms we associate vector fields on $G$ by
considering ker$d\theta^L_x$ and ker$d\theta^R_x$. These vector
fields close on the Lie algebra $\calG_x$, in terms of right or
left infinitesimal generators. We get two quotient spaces
$G/G_x^L$ and $G/G_x^R$, respectively. These two spaces carry
symplectic structures, namely the projections of $d\theta^L_x$ and
$d\theta^R_x$, respectively. These quotient spaces are
symplectomorphic with the symplectic orbit in $\calG^*$ passing
through $x\in\calG^*$.

It is this correspondence which allows us to compare the orbit
method with the construction of unitary representations via
functions on homogeneous spaces. We shall discuss it again by
using the reduction procedure on $T^*G$.

\section{Poisson subalgebras on $T^*G$ and associated quotient spaces}

The left invariant and right invariant splitting of $T^*G$, namely
$G{\times}_L\calG^*$ and $G{\times}_R\calG^*$, define a left invariant momentum
map projection $\calP^L: T^*G\rightarrow \calG^*$ and a right invariant
momentum map projection $\calP^R: T^*G\rightarrow \calG^*$. These maps are
Poisson maps with respect to the standard Poisson bracket on $T^*G$.
Therefore $\calF^L(\calG^*)\equiv (\calP^L)^*(\calF(\calG^*))$ and
$\calF^R(\calG^*)\equiv (\calP^R)^*(\calF(\calG^*))$ are Poisson subalgebras
of $\calF(T^*G)$ with the additional property
\begin{equation}
\{\calF^L(\calG^*),\calF^R(\calG^*)\}=0\,,
\end{equation}

\ni i.e. left and right momenta define mutually commuting Poisson subalgebras
in $\calF(T^*G)$. By considering these two projections we have the following
diagram:
\begin{eqnarray}
& G^L & \nn \\
&\downarrow& \ {}_{\calP^R} \nn \\
G^R \rightarrow\, &T^*G& \, \rightarrow \calG^*\,\,, \\
&\downarrow& \!\!\!\!{\scriptsize {\calP^L}} \nn\\
&\calG^*& \nn
\end{eqnarray}

\ni where $G^L$ and $G^R$ stay to represent a typical fiber of the projections
$\calP^L$ and $\calP^R$, respectively. It should be noticed that from the
point of view of groups we have a split sequence of groups
\begin{equation}
0\rightarrow\calG^*\rightarrow T^*G\rightarrow G^L\rightarrow 1 \,,
\end{equation}

\ni where $\calG^*$ is considered an abelian (vector) group.

We can use now the Marsden-Weinstein reduction procedure in a
revised form. We consider a point $m\in T^*G$ along with orbits
$G^L{\cdot} m$ and $G^R{\cdot} m$. By restricting $\w_0$ (the
canonical symplectic structure on $T^*G$) to this orbit we find
$\w^L_m=\w_0|_{G^L{\cdot} m}$ and $\w^R_m=\w_0|_{G^R{\cdot} m}$,
it is not difficult to show that $\w^L_m=d\theta^L_{\calP^L(m)}$
and $\w^R_m=d\theta^R_{\calP^R(m)}$, where $\calP^{L,R}(m)$ stay
for the corresponding point $x$ in $\calG^*$, according to the
appropriate projection. The kernel of $\w^{L,R}_m$  are generated
by vector fields associated with Casimir functions (or with
invariant relations, as the case may be) restricted to
$G^{L,R}{\cdot} m$, respectively. When projected onto $G$ they are
represented by vector fields on $G$ corresponding to $G_x$,
$x=\calP^{L,R}(m)$. With $G^{L,R}{\cdot} m$ we associate two
manifolds, $(G^L{\cdot} m)\cup (G^R{\cdot} m)$ and $(G^L{\cdot}
m)\cap (G^R{\cdot} m)$. The union of the two orbits is simply the
level set through $m$ of Casimir functions from $\calG^*$ to
$T^*G$. The intersection is the orbit (possibly a union of orbits)
of $G^L_x$ or $G^R_x$ with $x=\calP^{L,R}(m)$. Therefore a
neighborhood of a generic $m$ in $T^*G$ decomposes into the
product of two copies of symplectic co-adjoint orbits passing
through $x$ (for a generic $x$) and a symplectic manifold
diffeomorphic with $T^*G_x$ (this is not a symplectomorphism with
the standard symplectic structure on $T^*G$). This decomposition
does not hold for neighborhoods of singular symplectic orbits in
$\calG^*$.

To have a feeling of the problems arising at a global level, we
mention a different construction of previous decomposition.

On $\calG^*$ we restrict ourselves to the open sub-manifold of all
generic orbits. The pull-back of this sub-manifold to $T^*G$
defines an open symplectic sub-manifold on it. Level sets of
Casimir functions define a regular foliation. On these leaves we
restrict $\w_0$ and find a presymplectic structure. The kernel of
this 2-form is generated by the hamiltonian vector fields
associated with Casimir functions. The quotient with respect to
this kernel is a symplectic manifold, product of symplectic
orbits. This makes clear that this product arises in a quotient
space and is not a sub-manifold of the original leaf. These
remarks should make clear why our previously mentioned local
decomposition cannot be global, even if we restrict to generic
co-adjoint orbits.

For instance, in the case of $T^*SU(2)$ the product would give
$S^2{\times} S^2{\times} T^*S^1$, which clearly is not
diffeomorphic to $S^3{\times} R^3\equiv T^*SU(2)$. The open
sub-manifold of generic co-adjoint orbits would give rise to the
open sub-manifold $S^3{\times} S^2{\times} R$, again not
diffeomorphic to the mentioned decomposition.

\section{Polarized co-adjoint orbits: The generalized Jordan-Schwinger map}

The appearance of co-adjoint orbits of $G$ in $\calG^*$ in the
decomposition of $T^*G$ or in the homogeneous spaces $G/G_x$,
shows that to use the standard canonical quantization procedure
would be useful to find canonical (Darboux) coordinates for each
co-adjoint orbit in $\calG^*$, in this way we could use the
standard $p\rightarrow -i\hbar \frac{\partial\ }{\partial x}$ and
$x\rightarrow \hat{x}$ correspondence for the quantization. We can
rephrase this procedure by looking for a Poisson map $S: \calG^*
\rightarrow T^*Q$ which would allow to realize the algebra of $G$
in terms of Poisson brackets on $T^*Q$, with the Poisson bracket
on $T^*Q$ not being necessarily the standard one, i.e. we can
afford using commuting positions (on Q) and non-commuting momenta
(like the commutations relations of gauge invariant momenta in the
presence of magnetic fields, see \cite{Duval}). In this
formulation it is clear that co-adjoint orbits should be
noncompact for them to be diffeomorphic with cotangent bundles
(i.e. they should satisfy Pukanzsky condition, see \cite{Duval}),
moreover to realize the Lie algebra $\calG$ in terms of Poisson
brackets on $T^*Q$, this should allow for an action (at least
local) of $G$ so that $S$ becomes a $G$-invariant map.


This Poisson map is a classical version of the Jordan-Schwinger map, where
Lie algebras are realized in terms of creation and annihilation operators,
constituting a complexification of $(x,-i\hbar \frac{\partial\ }{\partial x})$.


We shall give more details of this construction, in order to apply it to
particular examples.

Let ${\cal O}_x$ be the co-adjoint orbit passing through the point
$x\in {\cal G}^*$. Let $\w$ be the symplectic form on ${\cal
O}_x$, obtained by inverting the Poisson brackets of $\calG^*$ on
$\calO_x$. If $\{p^1,p^2,\ldots,p^n\}$ is a basis\footnote{We
shall use the same notation $p^i$ for the momentum coordinates in
$T^*G$ and the coordinates in $\calG^*$.} of ${\cal G}^*$ (dual to
some selected basis in $\calG$), we can write $\w$ as:
\begin{equation}
\w = w_{ij}(p) dp^i\wedge dp^j\,.
\end{equation}

Let $S: \calG^*\rightarrow T^*Q$ be a Poisson map which allows to
write $\w$ in terms of Darboux coordinates
$q^i=q^i(p),\pi_i=\pi_i(p),\ i=1,\ldots,k$, $2k$ being the
dimension of the co-adjoint orbit ${\cal O}_x$, $\w = \sum_{i=1}^k
d\pi_i\wedge dq^i$. Then we define the following functions on
$T^*G$:
\begin{eqnarray}
X^i &\equiv& (\calP^L)^*{\cdot} S^*(q^i) \nn\\
P_{X^i} &\equiv& (\calP^L)^*{\cdot} S^*(\pi_i) \label{eq1}\\
& & \nn \\
Y^i &\equiv& (\calP^R)^*{\cdot} S^* (q^i) \nn \\
P_{Y^i} &\equiv& (\calP^R)^*{\cdot} S^*(\pi_i) \label{eq2} \,,
\end{eqnarray}

\ni for $i=1,2,\ldots,k$, where $\calP^{L,R}:T^*G\rightarrow \calG^*$ are the
generating functions
of the right (left) action of the group G on $T^*G$ (momentum maps). Then,
by definition,
$X^i,P_{X^i}$ (resp. $Y^i,P_{Y^i}$) are invariant under the left (resp. right)
action of the group, and since $\calP^{L,R}$ are Poisson maps, they verify:
\begin{eqnarray}
\{P_{X^i},X^j\}&=&\delta_{i}^j \nn\\
\{X^i,X^j\}&=&0 \nn\\
\{P_{X^i},P_{X^j}\}&=&0 \nn\\
& &  \\
\{P_{Y^i},Y^j\}&=&\delta_{i}^j \nn\\
\{Y^i,Y^j\}&=&0 \nn\\
\{P_{Y^i},P_{Y^j}\}&=&0 \nn \,.
\end{eqnarray}

As discussed above, we can generalize this situation to the case
in which the commutator between the $\pi_i$ among themselves do
not commute, to allow, for instance, for the description of gauge
invariant momenta in the presence of magnetic fields. In this case
the commutator between the $P_{X^i}$ (and the commutator between
the $P_{Y^i}$) will be different from zero.

Also, due to the fact that the left and the right action commute, both
set of coordinates, $(X^i,P_{X^i})$ and $(Y^i,P_{Y^i})$, commute (in other
words, since $(X^i,P_{X^i})$ are functions of the left invariant momenta
$(\calP^L)^i$ and $(Y^i,P_{Y^i})$ are functions of the right invariant
momenta $(\calP^R)^i$, they commute).

Note that we can use two different sets $(q^i,\pi_i)$ and $(q'^i,\pi'_i)$
for defining the coordinates $(X^i,P_{X^i})$ and $(Y^i,P_{Y^i})$,
respectively, and these would be related to the ones given before by a
canonical transformation in $T^*G$, since the two pairs of Darboux
coordinates are related through a canonical transformation in $T^*Q$.


Casimirs are functions on $\calG^*$ which are central with respect
to the natural Poisson bracket in $\calG^*$ (which is isomorphic
to the Lie algebra bracket of $\calG$). For every Lie algebra
there are a number of independent Casimirs (generally polynomials
in $\calG^*$) that is equal to the rank $r$ of the algebra. Taking
the pullback with $\calP^{L,R}$ of the Casimirs we obtain $r$
functions on $T^*G$ which are right and left invariant (and
therefore central with respect to the natural Poisson bracket in
$T^*G$). We define $P_{Z^i},\ i=1,\ldots,r$, as $r$ independent
functions of the Casimirs in such a way that, inverting the
equations (\ref{eq1},\ref{eq2}), $(\calP^L)^i$ could be written as
linear functions in the momenta $P_{X^j}$, and $(\calP^R)^i$ could
be linear in the momenta $P_{Y^j}$. Clearly, this will not be
always possible since it would imply that all co-adjoint orbits
for any group are polarizable (in the sense of Geometric
Quantization, see \cite{Woodhouse}), as we shall see in the next
section. But in the cases in which this construction is possible,
it simplifies very much the construction of unitary irreducible
representations (see next section).

 In some cases, we shall get only linearity in the momenta $P_{X^j}$ (resp.
$P_{Y^j}$) but not in the Casimirs $P_{Z^j}$. This problem is harmless
for our purposes, since it is still possible to define uniquely the
quantization (irreducible representations).

To complete the discussion, we only need to compute $r$ functions $Z^i$ on
$T^*G$ which are canonically conjugated to the $P_{Z^i}$ and such that
$(g^i,p^j) \rightarrow (X^i,Y^j,Z^k,P_{X^i},P_{Y^j},P_{Z^k})$ be a canonical
transformation in $T^*G$. These functions are neither left nor right invariant,
instead they transform according to an additive function on the group satisfying
the properties of 1-cocycle on the group (see \cite{Manko}).

\section{Irreducible representations}
\label{Irreducible}

Once we have found the canonical transformation which ``separates"
the dependence of the right and left momentum maps, it is very
easy to obtain irreducible representations of $G$ using the
techniques of Geometric Quantization in its simpler version. First
of all, note that from the construction realized above, we can
deduce few simple facts which will be useful in the following:
\begin{itemize}
\item The variables $P_{Z^i}$, being functions of the Casimirs in $T^*G$, are
invariant under the left and right action of $G$.
\item The variables $Z^i$, conjugated to $P_{Z^i}$, are cyclic in all
the generators of either the right or the left action. This implies that the
vector fields $\parcial{Z^i}$ commute with the generators of the right and left
action of $G$. Therefore, they must be constant on each irreducible
representation.
\end{itemize}

Our aim is to construct irreducible representations of $G$ using
the previous considerations in $T^*G$. We can proceed in two
different ways, one starting directly from a complex line bundle
on the phase-space $T^*G$ (the connection 1-form given by
$\Theta_0$, the curvature being the symplectic 2-form $d\Theta_0$
and the invariant measure given by $\Omega_0=(d\Theta_0)^n$) and
impose the appropriate polarization (and constraints) conditions
to obtain, finally, the desired representations of $G$. The second
possibility is to start with a certain subspace of $T^*G$, and
introduce on it a contact structure derived from the symplectic
structure in $T^*G$, and an invariant measure derived from the
natural volume in $T^*G$. We shall concentrate on the second
possibility since the first one requires more elaborated
procedures, involving (second class) constraints.

Consider the left\footnote{For concreteness, we shall consider one
of the two possible actions of the group, and construct our
procedure in terms of this action. Obviously, there exist the
analogous construction which leads to completely equivalent
results.} action of $G$ on $T^*G$. To this action, as usual, we
can associate a momentum map $\calP^R:T^*G\rightarrow \calG^*$. If
we consider a point\footnote{For the moment we shall restrict
ourselves to regular points in $\calG^*$.} $p$ in $\calG^*$, we
can define a hypersurface $\Sigma^R\subset T^*G$ as
$\Sigma^R\equiv {(\calP^R)}^{-1}(p)$, i.e. the level set of
$\calP^R$ associated with $p\in\calG^*$.

 On $\Sigma^R$ we have that $d\calP^R =0$, and from this
fact and the independence of the components of $\calP^R$ in $\calG^*$ we
deduce that
$d\Theta_0= dP_{X^i}\wedge dX^i$ when restricted to $\Sigma^R$
(remember that, in $T^*G$, $\Theta_0$ could be written as
$\Theta_0=P_{X^i} dX^i + P_{Y^i}dY^i + P_{Z^j}dZ^j$).

 Therefore, we can consider $\Sigma^R$ a contact manifold with presymplectic
form $d\Theta_0|_{\Sigma^R}$. The kernel of this presympectic
2-form is generated by the vector fields $\parcial{Z^j}$, and if
we quotient with the distribution generated by them we obtain a
symplectic manifold $\calS$ to which we can apply the techniques
of Geometric Quantization. The symplectic 2-form is
$\omega=d\Theta_0|_{\calS}=dP_{X^i}\wedge dX^i$ and there exist an
invariant volume (under the left action of the group) on this
symplectic manifold, which is clearly given by $\Omega\equiv
\omega^k=dP_{X^1}\wedge dX^1\wedge\ldots \wedge dP_{X^k}\wedge
dX^k$, where $2k$ is the dimension of the symplectic manifold
$\calS$. This symplectic manifold is symplectomorphic to the
co-adjoint orbit $\calO_p$ passing through the point $p\in\calG^*$
under the co-adjoint action of $G$.

The quantization of the symplectic manifold $\calS$ would proceed
as follows (see \cite{Woodhouse} for instance). Consider the
complex hermitian line bundle ${\cal L}$ on $\calS$ with
connection $\nabla$ and curvature $\omega=dP_{X^i}\wedge dX^i$.
The invariant (under the left action) volume is given by $\Omega$
and let us choose $\theta=P_{X^i}dX^i$ as the potential 1-form.
The Hilbert space $\calH$ is given by (the completion of) the
space of smooth sections on $\calL$ with scalar product given by
the volume $\Omega$. The ``quantum operator'' $\hat{f}$ associated
with the ``classical'' function $f$ on $\calS$ is defined in the
usual way, in terms of the lift of the vector field $V_f$
associated with $f$:
\begin{equation}
\hat{f}=-i \nabla_{V_f} + f I\,,
\end{equation}

\ni where $V_f\equiv\{f,{\cdot}\}=\frac{\partial f}{\partial
P_{X^i}}\parcial{X^i}- \frac{\partial f}{\partial
X^i}\parcial{P_{X^i}}$ and $I$ is the identity operator. If
$\calS$ is a cotangent bundle\footnote{This will always be our
case by construction, since we are assuming that we can find
(global) Darboux coordinates in the co-adjoint orbit diffeomorphic
to $\calS$.}, the connection is written like $\nabla_V\Psi=V\Psi
-i \theta(V)\Psi$, for any vector field $V$ in $\calS$.

 The left-invariant momenta $\calP^L_i$, which can be written in terms of
the left-invariant variables $X^j,\ P_{X^j}$ and the Casimirs
$P_{Z^k}$, restrict to the hypersurface $\Sigma^R$, and also to
the symplectic manifold $\calS$, i.e. they continue to be the
generating functions of the (right) action of $G$ on $\Sigma^R$
and $\calS$ since the Casimirs $P_{Z^k}$ are constant on
$\Sigma^R$ and $\calS$. The quantum operators $\hat{\calP}^L_i$,
obtained as the lift of the associated vector fields on $\calS$,
provide us with a representation of the group $G$ acting on
$\calH$. This representation is unitary with the scalar product
given by the invariant measure $\Omega$. As it is well-known, this
representation is not irreducible and some invariant subspace
$\calH'\subset\calH$ should be selected. This task is achieved by
means of the polarization conditions, expressed in terms of
operators acting on $\calH$ obtained as the ``horizontal lift'' of
vector fields associated with certain functions on $\calS$, which
is given simply by $\tilde{f}=\nabla_{V_f}$.
%

 The polarization conditions are imposed through a Poisson subalgebra $\calP$
of the Poisson algebra of functions on $\calS$. This Poisson
subalgebra $\calP$ defines a Hilbert subspace $\calH'\subset
\calH$ on which the ``horizontal lift'' of all vector fields
associated with functions in $\calP$ vanish (in other words,
$\calH'$ is the subspace of sections which are ``parallel'' to the
integral curves of all vector fields associated to functions in
$\calP$). Clearly, for this condition to be compatible with the
representation of $G$, i.e.  for $\calH'$ to be an invariant
subspace under the action of $G$, it is necessary that the
condition $\{\calP^L_i,\calP\}\subset \calP$ be satisfied. In this
case we shall say that the polarization subalgebra $\calP$ is {\it
admissible}. This condition imposes strong restrictions on the
choices (and even the existence) of polarization subalgebras
$\calP$ for each given group $G$.


Also, if the left invariant momenta are polynomials of degree up
to one in the momenta $P_{X^i}$, we find easily an admissible
polarization subalgebra, since the abelian subalgebra $\calP^X$ of
functions of $X^i$ is always preserved  by the momentum maps
$\calP^L_j$. Therefore, imposing the restrictions:
\begin{equation}
\tilde{X}^i\ \Psi=\nabla_{V_{X^i}}\Psi = -\parcial{P_{X^i}}\Psi =0\, ,
\end{equation}

\ni we select a subspace $\calH^X$ of wave functions in $\calH$
that depend only on the coordinates (coordinate representation)
and such that it is invariant under the action of $G$. This
subspace will provide, in most of the cases, an irreducible
representation of $G$ with measure (not necessarily invariant)
$dX^1\wedge\ldots\wedge dX^k$. To guarantee the irreducibility, we
should also take into account the possible existence of
higher-order differential operators (see
\cite{Higher-Order,Bregenz}) or even discrete operators commuting
with the representation, and which escapes to our local treatment.
In the case in which the measure is not invariant, it will be
quasi-invariant and therefore the representation can be unitarized
by means of the Radon-Nikodym derivative (see \cite{Barut}).

 Depending on the form of the left-invariant momenta, there can exist more
admissible polarization subalgebras leading to equivalent or
inequivalent representations, this may change from case to case.
If they are polynomials of degree up to one in the coordinates
$X^i$, then the abelian subalgebra $\calP^{P_X}$ of functions of
$P_{X^i}$ defines an admissible polarization subalgebra.
Therefore, imposing the restrictions:
\begin{equation}
\tilde{P}_{X^i}\ \Psi = \left(\parcial{X^i}-iP_{X^i}\right)\Psi =0\, ,
\end{equation}

\ni we select a subspace $\calH^{P_X}$ of wave functions depending,
up to a phase factor, on the momenta (momentum representation), which is
also invariant under the action of $G$.

It is worth to note that (in the case in which both polarizations
are admissible) they are related by the ``adjoint'' action of a
certain function in $\calS$. This is nothing other than (the
generator of) the Fourier transform, associated with the function
$\calF\equiv \medio \sum_{i=1}^k(X^i{}^2+P_{X^i}^2)$, verifying
$\{\calF,\calP^X\}=\calP^{P_X}$ and
$\{\calF,\calP^{P_X}\}=\calP^{X}$.
The coordinate and momentum representations are unitarily
equivalent, since the Fourier transform is a unitary operator.
However, the Fourier transform is an integral operator and cannot
be obtained within the framework of Geometric Quantization (one
has to resort to the Metaplectic representation, see for instance
\cite{Folland}). The generator of the Fourier transform can be
obtained as a second order differential operator using the
technique of higher-order polarizations
\cite{Higher-Order,Bregenz} (see below).

 It is interesting to note that, even in the case in which only one of the
polarization subalgebras $\calP^X$ or $\calP^{P_X}$ is admissible,
this does not mean that the other representation does not exist.
In fact, one can apply the Fourier transformation and obtain an
equivalent representation. The difference arises in the fact that
in this representation the operators $\hat{\calP}^L_i$ will be
higher-order differential operators, which cannot be obtained
within the framework of Geometric Quantization. Using other
techniques, like the technique of higher-order polarizations (see
\cite{anomalias,Higher-Order,Bregenz}, for instance), one can
obtain them, even in the anomalous case in which none of the
polarizations exist, as it happens for the Schr\"{o}dinger group.

Essentially, the technique of higher-order polarizations consists
in allowing to enter  the polarization subalgebra higher-order
differential operators, which are available in the enveloping
algebra of the Lie algebra. More precisely, a higher-order
polarization is a maximal subalgebra of the (left or right)
universal enveloping algebra with no intersection with the
identity operator on sections. With this definition, a
higher-order polarization contains the maximum number of
conditions compatible with the hermitian line bundle structure of
${\cal L}$ and with the action of the quantum operators.

The use of higher-order polarizations allows, for instance, to obtain the
Schr\"{o}dinger equation in configuration space, which is a second order
differential equation, or to find a polarization for the anomalous case of
the Schr\"{o}dinger group, which does not admit any usual, first order,
polarization. However, in this work we shall restrict to usual, first order
polarizations.


 Finally, there exist polarization conditions which are not related to
polarization subalgebras, i.e. they are not the ``horizontal lift'' of any
vector field in $\calS$, but define a subspace $\calH'\subset \calH$ that
is left invariant by the action of $G$. We shall see an example of such
polarization condition for the case of the 2-dimensional group.

\section{Examples}

 In this section we shall apply the results of previous sections to some
particular examples, the simplest case of the non-trivial two dimensional Lie
group, and all
three dimensional Lie groups.

For all these groups, generic co-adjoint orbits have dimension 2,
for which Darboux coordinates $p,q$ can be found and their inverse
images by the (left and right invariant) momentum maps provide us
with the coordinates (left and right invariant, respectively)
$X,P_X$ and $Y,P_Y$. Choosing an arbitrary point $p$ in each
co-adjoint orbit, we construct the hypersurface $\Sigma^R$ as
explained in the previous section, which is parametrized by
$X,P_X$ and $Z$ (in the case of the 2-dimensional group the
variable $Z$ is not present since there is no Casimir). The
presymplectic 2-form $d\Theta_0|_{\Sigma^R}$ is $dP_X\wedge dX$,
which has $\parcial{Z}$ as kernel (except for the 2-dimensional
group, for which it is directly symplectic). Taking quotient by
the distribution generated by this kernel we obtain a symplectic
manifold $\calS$ with symplectic 2-form
$\omega=d\Theta_0|_{\calS}=dP_X\wedge dX$. Although apparently the
construction is identical for all the groups (this is because we
are working with Darboux coordinates), $\calS$ differs for each
group, since it is a homogeneous manifold with respect to its
corresponding group, and the range of the parameters $X$ and $P_X$
can also be different for each group.

 The construction of the hermitian line bundle $\calL$ is also similar in all
 cases, with the connection $\nabla$ given by the potential
1-form $\theta=P_XdX$ and curvature $\omega=dP_X\wedge dX$. The
Hilbert space $\calH$ of smooth sections on $\calL$ completed with
respect to the scalar product given by the invariant measure
$\Omega=\omega$ is also constructed along the guidelines of
previous section. The ``vertical lift'' of a vector field on
$\calS$ to a (quantum) operator on $\calH$ is constructed as
explained previously, and the same for the construction of
polarization conditions by means of the ``horizontal lift'' of
vector fields. The differences for each groups lies in the
explicit form of the operators $\hat{\calP}^L$ associated with the
left-invariant momentum maps of the action of $G$ on $\cal$ and in
the existence or not of certain polarization subalgebras $\calP$.


\subsection{The 2-dim group}

 Consider the group of matrices

\begin{equation}
\left(
\begin{array}{ll}
x & y \\
0 & 1
\end{array}
\right) \,,
\end{equation}
with $x\in R-\{0\}$ and $y\in R$. This group can be identified with the group
of scale transformations and translations on the real line:
\[
u^{^{\prime }}=xu+y
\]
It is not connected unless $x$ is restricted to be positive, $G^{+}$ will
denote the component of $G$ connected to the identity. This group is
solvable and therefore unitary representations are either one dimensional or
infinite dimensional.

It is also diffeomorphic to $SB(2,R)$, the diffeomorphism being
\[
\left(
\begin{array}{ll}
x & y \\
0 & \frac{1}{x}
\end{array}
\right) \Longleftrightarrow \left(
\begin{array}{ll}
\frac{x^{3}}{\left| x\right| } & xy \\
0 & 1
\end{array}
\right)
\]

>From matrix multiplication we can obtain easily the group law:
\begin{eqnarray}
x'' &=& x'x \nn \\ y'' &=& y'+x'y \,,
\end{eqnarray}

\ni and from this right and left invariant vector fields are easily computed:
\begin{equation}
\begin{array}{rcl}
X^L_x &=& x\parcial{x} \\
X^L_y &=& x\parcial{y}
\end{array}\hbox{\hskip 1cm}
\begin{array}{rcl}
X^R_x &=& -x\parcial{x}-y\parcial{y} \\
X^R_y &=& -\parcial{y}\,.
\end{array}
\end{equation}

The Lie algebra for this group is given by:
\[
\left[ X_{x}^{L},X_{y}^{L}\right] =X_{y}^{L} \,.
\]

This Lie algebra, being two dimensional and non-trivial, admits no Casimirs.
The left and right-invariant momentum maps
$\calP^{L,R}:T^*G^+\rightarrow \calG^*$ of the action of
$G^+$ on $T^*G^+$ are computed, as usual, by
$\calP^{L,R}_i\equiv i_{X^{L,R}_i}\Theta_0$, where $\Theta_0$ is the
canonical 1-form on $T^*G^+$, $\Theta_0= p_x dx + p_y dy$.
We obtain:
\begin{equation}
\begin{array}{rcl}
\calP^L_x &=& xp_{x} \\
\calP^L_y &=& xp_{y}
\end{array}\hbox{\hskip 1cm}
\begin{array}{rcl}
\calP^R_x &=& -xp_{x}-yp_y \\
\calP^R_y &=& -p_{y}
\end{array}
\end{equation}

To find the canonical variables $(X,Y,P_X,P_Y)$ in $T^*G^+$, we
proceed as explained before, finding Darboux coordinates on
co-adjoint orbits. In this case, co-adjoint orbits by the action
of $G^+$ are given by two half-planes $A^+=\{p_y>0\}$ and
$A^-=\{p_y<0\}$ and the zero-dimensional orbits
$A^0_{p_x}=\{p_x,p_y=0\}$. If the non-connected group $G$ is
considered, $A^+\cup A^-$ is a single two-dimensional disconnected
orbit.

In the dual space of the Lie algebra there is a natural Poisson bracket
\[
\Lambda =p_{y}\frac{\partial }{\partial p_{x}}\wedge \frac{\partial }{\partial
p_{y}}\,,
\]
which is non degenerate on $A^{+}$ and $A^{-}$ . On $A^{+}$ and $A^{-},$
$\Lambda $ has an inverse
\[
\w =\left( \Lambda \right) ^{-1}=\frac{1}{p_{y}}dp_{x}\wedge dp_{y}
\]

A set of Darboux coordinates for $\w$ is given by the map
$S:\calG^*\rightarrow T^*R$:
\begin{equation}
S(p_x,p_y)=(p_y,\frac{p_x}{p_y})\equiv (\pi,q)
\end{equation}

Making use of the left and right-invariant momentum maps, we obtain:
\begin{eqnarray}
X &=& (\calP^L)^*{\cdot} S^*(q)=\frac{\calP^L_x}{\calP^L_y}=\frac{p_x}{p_y}\nn \\
P_X &=& (\calP^L)^*{\cdot} S^*(\pi)=\calP^L_y = xp_y \nn \\
 & & \\
Y&=& (\calP^R)^*{\cdot} S{}^*(q)=\frac{\calP^R_x}{\calP^R_y} =
 x \frac{p_x}{p_y} + y \nn \\
P_Y&=& (\calP^R)^*{\cdot} S{}^*(\pi)=\calP^R_y = - p_y
\end{eqnarray}

Inverting these relations we get the expressions of $\calP^{L,R}_i$
in terms of $X,P_X$ (resp. $Y,P_Y$):
\begin{eqnarray}
\calP^L_x &=& XP_X \nn\\
\calP^L_y &=& P_X\nn\\
 & & \label{2-dimirred}\\
\calP^R_x &=& Y P_Y \nn\\
\calP^R_y &=& P_Y \,. \nn\\
\end{eqnarray}

Now we proceed with the computation of the irreducible representations. We
shall consider only the 2-dimensional orbits in $\calG^*$, $A^+$ and $A^-$,
which are generic. The non-generic ones are only, in this case, the
zero-dimensional orbits. These orbits are associated with
one-dimensional representations of $G^+$, i.e. characters of $G^+$, and can be
computed easily with other procedures. Also, the representations associated
to $A^+$ and $A^-$ are equivalent, since the discrete operator of $G$
which sends one orbit to the other is unitary.

Thus, let us consider a point $p\in A^+$, for instance $p=(0,p_y^0>0)$.
The hypersurface $\Sigma^R$ associated with this point is given by
$\Sigma^R\equiv \calP^R{}^{-1}(p)=\{(x,y,p_x,p_y)\in T^*G^+,\
{\rm such\  that\ } xp_x+yp_y=0,\ p_y=p_y^0\}$. The variables $Y$ and $P_Y$ are
constant on $\Sigma^R$, which is parameterized by $X=-y/x$ and $P_X=xp_y^0>0$.

In this particular case the presymplectic 2-form $d\Theta_0|_{\Sigma^R}$
has no kernel and therefore
the symplectic manifold $\calS$ coincides with $\Sigma^R$ and $\omega$ with
$d\Theta_0|_{\Sigma^R}$.

The vertical lift of the vector fields associated with $\calP^L_x$
and $\calP^L_y$ are:
\begin{equation}
\hat{\calP}^L_x = -iX\parcial{X} +i P_X\parcial{P_X}\,,\qquad
\hat{\calP}^L_y = -i\parcial{X}\, .
\end{equation}

Since the left invariant momenta $\calP^L_x,\calP^L_y$ are both
polynomials of degree up to one in $P_X$ and $X$, the polarization
subalgebras $\calP^X$ and $\calP^{P_X}$ are both admissible, and
lead to equivalent representations related by the Fourier
transform. Using $\calP^X$, for instance, the restriction that we
have to impose is given by $\tilde{X}\Psi = -\parcial{P_X}\Psi
=0$, therefore obtaining a Hilbert subspace $\calH^X$ of sections
depending only on $X$ on which the operators $\hat{\calP}^L_x$ and
$\hat{\calP}^L_y$ have the form:
\begin{eqnarray}
\hat{\calP}^L_x\Psi(X) &=& -iX\parcial{X} \Psi(X) \nn \\
\hat{\calP}^L_y\Psi(X) &=& -i\parcial{X} \Psi(X)\, .
\end{eqnarray}

\ni This clearly constitute an irreducible representation of the 2-dimensional
group. It is not unitary since the measure $dX$ on $\calH^X$ is not invariant,
but quasi-invariant.
Introducing the Radon-Nykodim derivative we obtain a unitary representation
in which $\hat{\calP}^L_x\Psi(X) =-i(X\parcial{X}+\medio) \Psi(X)$, as it
should be \cite{Barut}.

There exist another polarization which does not come from any polarizing
subalgebra $\calP$ in $\calS$. This is given by the condition:
\begin{equation}
\parcial{X}\Psi = 0\,,
\end{equation}

\ni which leads to sections depending only on $P_X$ (do not confuse with the
ones
given by the polarization $\calP^{P_X}$, which carry an additional phase of
the form $e^{iXP_X}$). This defines an invariant Hilbert space
$\calH^{P_X}{}'$ on which the operators act as:
\begin{eqnarray}
\hat{\calP}^L_x\Psi(P_X) &=& iP_X\parcial{P_X} \Psi(P_X) \nn \\
\hat{\calP}^L_y\Psi(P_X) &=& 0 .
\end{eqnarray}

This representation, with measure $dP_X$ and corrected with the
Radon-Nikodym deriva\-tive\footnote{In this particular case it is
already unitary with respect to the invariant measure $dP_X/P_X$,
since the translations are represented trivially.}, is clearly non
equivalent to the previous one, since one is faithful and the
other one is not (the operator $\hat{\calP}^L_y$ is represented
trivially).

\subsection{The Heisenberg-Weyl group}

In quantum optics, the Heisenberg--Weyl group plays a very
important role. The quadrature components of photon are the
generator of Lie algebra of the Heisenberg--Weyl group. The photon
creation and annihilation operators are linear combinations of the
Lie algebra generators.

The Heisenberg-Weyl group in one dimension is a non-trivial central
extension of $R^2$ by $R$ or $U(1)$. Its elements can be written in matrix form:
\begin{equation}
g=\left( \begin{array}{ccc} 1 & x & z \\
0 & 1& y \\
0 & 0 & 1
\end{array} \right) \,.
\end{equation}

The group law $g''=g'*g$ for this group is given by:
\begin{eqnarray}
x'' &=& x' + x \nn \\
y'' &=& y'+ y \\
z'' &=& z' + z + x'y \nn \,.
\end{eqnarray}

Left and right-invariant vector fields are given by:
\begin{equation}
\begin{array}{rcl}
X^L_x &=& \parcial{x} \\
X^L_y &=& \parcial{y}+x\parcial{z} \\
X^L_z&=&\parcial{z}
\end{array}\hbox{\hskip 1cm}
\begin{array}{rcl}
X^R_x &=& -\parcial{x}-y\parcial{z} \\
X^R_y &=& -\parcial{y} \\
X^R_z&=&-\parcial{z} \,.
\end{array}
\end{equation}

The commutation relations are:
\begin{equation}
[X^L_x,X^L_y ] = X^L_z \label{LieH-W}
\end{equation}

This Lie algebra admits a Casimir, which is given by the central element of
the Lie algebra $X^L_z=-X^R_z=\parcial{z}$.

We compute the left and right-invariant momentum maps
$\calP^{L,R}:T^*G\rightarrow \calG^*$ as
usual, $\calP^{L,R}_i\equiv i_{X^{L,R}_i}\Theta_0$, where $\Theta_0$ is the
canonical 1-form on $T^*G$, $\Theta_0= p_\theta d\theta + p_x dx + p_y dy$.
We obtain:
\begin{equation}
\begin{array}{rcl}
\calP^L_x &=& p_{x} \\
\calP^L_y &=& p_{y} + xp_{z}\\
\calP^L_z &=& p_z
\end{array}\hbox{\hskip 1cm}
\begin{array}{rcl}
\calP^R_x &=& -p_{x}-yp_z \\
\calP^R_y &=& -p_{y}\\
\calP^R_\theta &=& -p_z \,.
\end{array}
\end{equation}

The Lie algebra, in terms of Poisson brackets, generated by $\calP^{L,R}_i$
is isomorphic to that of $\calG$ in (\ref{LieH-W}).

The classical Casimir is given by $p_z$. To find the canonical
variables $(X,Y,Z,P_X,P_Y,P_Z)$ in $T^*G$ we proceed as explained
before, computing Darboux coordinates on co-adjoints orbits. For
the H-W group, the co-adjoints orbits are of two classes:
two-dimensional orbits of the form ${\cal O}_{p_z^0}=\{
(p_x,p_y,p_z), p_x,p_y\in R \}$, with $p_z=p_z^0\neq 0$, and
zero-dimensional orbits of the form ${\cal
O}_{p_x,p_y}=\{(p_x,p_y,0)\}$. Obviously, we are interested in the
two-dimensional orbits ${\cal O}_{p_z^0}$.

 The symplectic form $\w$ for $\calO_{p_z^0}$, obtained by inverting the
Poisson brackets on the orbit, has the form:
\begin{equation}
\w=\frac{1}{p_z^0}dp_x\wedge dp_y\,,
\end{equation}

\ni (remember that $\{p_x,p_y\}=p_z$ and $p_z$ is central). A set of Darboux
coordinates is given, for example, by the map $S: \calG^*\rightarrow T^*R$
with $S(p_x,p_y,p_z)=(\frac{p_y}{p_z^0},p_x)=(q,\pi)$ or by the map
$S': \calG^*\rightarrow T^*R$ with
$S'(p_x,p_y,p_z^0)=(\frac{p_x}{p_z},-p_y)=(q',\pi')$. Therefore,
making use of the left and right momentum maps, we obtain:
\begin{eqnarray}
X &=& (\calP^L)^*{\cdot} S^*(q)=\frac{\calP^L_y}{\calP^L_z}=
        x+\frac{p_y}{p_z^0}\nn \\
P_X &=& (\calP^L)^*{\cdot} S^*(\pi)=\calP^L_x=p_x \nn\\
& &  \\
Y &=& (\calP^R)^*{\cdot} S'{}^*(q')=\frac{\calP^R_x}{\calP^R_z}=
        y+\frac{p_x}{p_z^0}\nn \\
P_Y &=& (\calP^R)^*{\cdot} S'{}^*(\pi')=\calP^R_y=p_y \nn \,.
\end{eqnarray}

Inverting these relations, we can obtain the expressions of $\calP^{L,R}_i$
in terms of $X,P_X,P_Z$ (resp. $Y,P_Y,P_Z$):
\begin{eqnarray}
\calP^L_x &=& P_X \nn\\
\calP^L_y &=& X P_Z \nn\\
\calP^L_z &=& P_Z \nn\\
 & & \label{H-Wirred}\\
\calP^R_x &=& Y P_Z \nn\\
\calP^R_y &=& P_Y \nn\\
\calP^R_z &=& P_Z \nn\,,
\end{eqnarray}

\ni where the Casimir $P_Z$ is easily computed to be $p_z$.
To compute the cyclic variable $Z$, we make use of the fact that
\begin{equation}
\Theta_0 = p_xdx+p_ydy+p_zdz = P_XdX+P_YdY + P_ZdZ\,,
\end{equation}

\ni from which we obtain directly that $Z=z-\frac{p_xp_y}{p_z^2}$.


For the computation of the irreducible representations, let us
consider, as explained in Sec. \ref{Irreducible}, a point $p$
belonging to a generic orbit in $\calG^*$. In this case the
generic orbits are the two dimensional orbits $\calO_{p_z^0}$,
with $p_z^0\neq 0$. We can choose, for simplicity,
$p=(0,0,p_z^0)$. This defines a hypersurface of the form
$\Sigma^R=\calP^L{}^{-1}(p)=\{(x,y,z,p_x,p_y,p_z)\in T^*G\ {\rm
such\ that\ } p_x+yp_z=0,\,p_y=0,\,p_z=p_z^0\}$ on $T^*G$, which
is parameterized, in this case, by the coordinates $X=x,\,P_X=p_x$
and $Z=z$.



The vertical lift of the vector fields associated with the left invariant
momenta are:
\begin{eqnarray}
\hat{\calP}^L_x &=& -i\parcial{X}\nn \\
\hat{\calP}^L_y &=& iP_Z\parcial{P_X}+ XP_ZI \label{H-Woperators}\\
\hat{\calP}^L_z &=& P_ZI \nn \,.
\end{eqnarray}

 Again, the left invariant momenta are all polynomials of degree up to one
in the variables $X$ and $P_X$, implying that both polarization subalgebras
$\calP^X$ and $\calP^{P_X}$ are admissible and lead to equivalent
representations, related by the Fourier Transform. Using, for instance,
$\calP^X$ we obtain the same Hilbert subspace $\calH^X$ of sections
depending only on the variable $X$ as in the case of the 2-dimensional group.
The action of the operators in (\ref{H-Woperators}) on this Hilbert space
is given by:
\begin{eqnarray}
\hat{\calP}^L_x\Psi(X) &=& -i\parcial{X} \Psi(X) \nn \\
\hat{\calP}^L_y\Psi(X) &=& XP_Z\Psi(X) \\
\hat{\calP}^L_z\Psi(X) &=& P_Z\Psi(X)\,. \nn
\end{eqnarray}

This constitutes a unitary and irreducible representation of the
H-W group, and coincides with the standard Schr\"{o}dinger
representation with the identification $P_Z\equiv \hbar$. Thus the
eigenvalue of the Casimir operator is identified with the Planck's
constant.

 Unlike the case of the 2-dimensional group, $\parcial{X}$ does not define
a polarization condition, since the subspace of sections depending
only on $P_X$ (without any additional phase as the one that
results from the polarization $\calP^{P_X}$) is not invariant
under the H-W group. Therefore these (one for each value of $P_Z$)
are the only unitary irreducible representations of the
Heisenberg-Weyl group, apart from the one-dimensional ones.

\subsection{The Euclidean group $E(2)$}

The Euclidean group in two dimensions is the semidirect action of
$U(1)$ on $R^2$, i.e. it is constituted by translations and
rotations in the plane. Its elements can be written in a matrix
form as:
\begin{equation}
g=\left( \begin{array}{ccc}
\cos\theta & \sin\theta & x \\
-\sin\theta & \cos\theta & y \\
0 & 0 & 1
\end{array} \right) \,.
\end{equation}

The group law $g''=g'*g$ for this group is given by:
\begin{eqnarray}
\theta'' &=& \theta' + \theta \nn \\
x'' &=& x' + \cos\theta' x + \sin\theta' y \\
y'' &=& y' -\sin\theta' x + \cos\theta' y \nn \,.
\end{eqnarray}

Left and right-invariant vector fields are given by:
\begin{equation}
\begin{array}{rcl}
X^L_\theta &=& \parcial{\theta} \\
X^L_x &=& \cos\theta\parcial{x} - \sin\theta\parcial{y} \\
X^L_y &=& \cos\theta\parcial{y} + \sin\theta\parcial{x}
\end{array}\hbox{\hskip 1cm}
\begin{array}{rcl}
X^R_\theta &=& -\parcial{\theta} +y\parcial{x}-x\parcial{y} \\
X^R_x &=& -\parcial{x} \\
X^R_y &=& -\parcial{y} \,.
\end{array}
\end{equation}

The commutation relations are:
\begin{equation}
[X^L_\theta,X^L_x ] = X^L_y \,,\,\,\,\,\,\,\,\, [X^L_\theta,X^L_y ]=- X^L_x\,.
\end{equation}

This Lie algebra admits a Casimir, which is given by the second order operator
\begin{equation}
\hat{C}_2 = (X^{L,R}_x)^2 + (X^{L,R}_y)^2\,.
\end{equation}

We compute the left and right-invariant momentum maps
$\calP^{L,R}:T^*G\rightarrow \calG^*$ as
usual, $\calP^{L,R}_i\equiv i_{X^{L,R}_i}\Theta_0$, where $\Theta_0$ is the
canonical 1-form on $T^*G$, $\Theta_0= p_\theta d\theta + p_x dx + p_y dy$.
We obtain:
\begin{equation}
\begin{array}{rcl}
\calP^L_\theta &=& p_\theta \\
\calP^L_x &=& \cos\theta p_{x} - \sin\theta p_{y} \\
\calP^L_y &=& \cos\theta p_{y} + \sin\theta p_{x}
\end{array}\hbox{\hskip 1cm}
\begin{array}{rcl}
\calP^R_\theta &=& -p_{\theta} +yp_{x}-xp_{y} \\
\calP^R_x &=& -p_{x} \\
\calP^R_y &=& -p_{y}\,.
\end{array}
\end{equation}

The classical Casimir is given by
$C_2=(\calP^{L,R}_x)^2 + (\calP^{L,R}_y)^2 = p_x^2+p_y^2$.

To find the canonical variables $(X,Y,Z,P_X,P_Y,P_Z)$ in $T^*G$ we
proceed as explained before, computing Darboux coordinates on
co-adjoints orbits. For the case of $E(2)$, co-adjoint orbits are
easily computed. The two-dimensional co-adjoints orbits $\calO_R$
correspond to the cylinders given by the equation
$C_2=p_x^2+p_y^2=R^2$, for $R>0$, and the zero-dimensional orbits
are given by $\calO_{p_\theta}=\{ (p_\theta,0,0)\}$. We are
interested in the two-dimensional orbits, for which the symplectic
form is (computed again by inverting the Poisson brackets on the
co-adjoint orbit):
\begin{equation}
\w = \frac{1}{R^2}dp_\theta\wedge (p_ydp_x - p_xdp_y)\,.
\end{equation}

A set of Darboux coordinates for $\w$ is given by the map
$S: \calG^*\rightarrow T^*S^1$:
\begin{equation}
S(p_\theta,p_x,p_y)=(p_\theta,\tan^{-1} \frac{p_x}{p_y})=(\pi,q)\,.
\end{equation}

Once we have found Darboux coordinates on the co-adjoint orbits,
we simple define $X,P_X$ (resp. $Y,P_Y$) as the pullback by
$\calP^L$ (resp. $\calP^R$) of $S^*(q)$ and $S^*(\pi)$,
respectively:
\begin{eqnarray}
X&=& (\calP^L)^*{\cdot} S^*(q) = \tan^{-1}\frac{\calP^L_x}{\calP^L_y}=
           \tan^{-1}\frac{\cos\theta p_x-\sin\theta p_y}{\cos\theta p_y+\sin\theta p_x}\nn \\
P_X &=& (\calP^L)^*{\cdot} S^*(\pi)=\calP^L_\theta = p_\theta \nn \\
 & & \\
Y&=& (\calP^R)^*{\cdot} S^*(q) = \tan^{-1}\frac{\calP^R_x}{\calP^R_y}=
           \tan^{-1}\frac{p_x}{p_y} \nn \\
P_Y &=& (\calP^R)^*{\cdot} S^*(\pi)=\calP^R_\theta = -p_\theta +yp_x -xp_y\nn \,.
\end{eqnarray}

\ni  Inverting these relations we obtain the expressions of $\calP^{L}_i$
in terms of $X,P_X$ and $P_Z$:
\begin{eqnarray}
\calP^L_\theta &=& P_X \nn \\
\calP^L_x &=& \sin X\ P_Z \\
\calP^L_y &=& \cos X\ P_Z \nn\,,
\end{eqnarray}

\ni with analogous expressions for $\calP^R_i$ in terms of $Y,P_Y$ and $P_Z$,
where $P_Z$ is the function of the Casimir $P_Z=R=\sqrt{C_2}$.
The only thing we have
to compute is the cyclic coordinate $Z$ canonically conjugated to $P_Z$. As in
the previous
example, using the fact that the canonical 1-form $\Theta_0$ in $T^*G$ can be
written as
$P_XdX+P_YdY+P_ZdZ$, we derive the expression for $Z$:
\begin{equation}
Z=\frac{xp_x+yp_y}{\sqrt{p_x^2+p_y^2}}\,.
\end{equation}


 Following Sec. \ref{Irreducible}, let us compute the irreducible representations. Choose a
point $p$ in each 2-dimensional co-adjoint orbit, characterized by
radius $R$, for instance $p=(0,R,0)$. The hypersurface $\Sigma^R$
is given by
$\Sigma=\calP^L{}^{-1}(p)=\{(x,y,\theta,p_x,p_y,p_\theta)\in T^*G\
{\rm such\ that\ } p_\theta-yp_x-xp_y=0,\,p_x=R,\,p_z=0\}$.
$\Sigma^R$ is then parametrized by the coordinates
$X=\frac{\pi}{2}-\theta$, $P_X=yR$ and $Z=x$.

The vertical lift of the momentum maps are given by:
\begin{eqnarray}
\hat{\calP}^L_\theta &=& -i\parcial{X}\nn \\
\hat{\calP}^L_x &=& i\cos X P_Z\parcial{P_X} +
                     \sin XP_ZI \label{E(2)-operators}\\
\hat{\calP}^L_y &=& -i\sin X P_Z\parcial{P_X} + \cos X P_ZI \nn \,.
\end{eqnarray}

In this case, the momentum maps are all up to first order in the momentum
$P_X$, therefore the polarization subalgebra $\calP^X$ is admissible, but are
not polynomial in $X$, implying that the polarization subalgebra $\calP^{P_X}$
is not admissible. Using $\calP^X$, we obtain the representation in coordinate
space, with sections depending only on $X$, and operators with the form:
\begin{eqnarray}
\hat{\calP}^L_\theta\Psi(X) &=& -i\parcial{X} \Psi(X) \nn \\
\hat{\calP}^L_x\Psi(X) &=& P_Z\sin X\Psi(X) \\
\hat{\calP}^L_y\Psi(X) &=& P_Z\cos X\Psi(X)\,. \nn
\end{eqnarray}

This representation is irreducible and unitary with respect to the measure
$dX$. Up to equivalence, it is the only representation for each value of
$P_Z=R$ of the group $E(2)$ (that is, there are no other polarization
conditions leading to inequivalent representations such as happened
in the 2-dimensional group with $\parcial{X}$).

\subsection{The $SB(2,C)$ group}

The $SB(2,C)$ group is given by the following matrix representation:
\begin{equation}
\left( \begin{array}{cc}
x & y+iz \\
0 & x^{-1}
\end{array} \right)\,,
\end{equation}

\ni with $x\in R-\{0\}$, and $y,z\in R$. As in the case of
$SB(2,R)$, it is a non-connected group, the connected component of
the identity $G^+$ is given by $x\in R^+$.

The group law is easily obtained from matrix multiplication, being:
\begin{eqnarray}
x'' &=& x'x \nn\\
y'' &=& \frac{y'}{x} + x'y \\
z''&=& \frac{z'}{x} + x'z\,. \nn
\end{eqnarray}

Left and right invariant vector fields are:
\begin{equation}
\begin{array}{rcl}
X^L_x &=& x\parcial{x} -y\parcial{y} -z\parcial{z} \\
X^L_y &=& x\parcial{y} \\
X^L_z &=& x\parcial{z}
\end{array} \qquad\qquad
\begin{array}{rcl}
X^R_x &=& -x\parcial{x} -y\parcial{y} -z\parcial{z} \\
X^R_y &=& -x^{-1}\parcial{y} \\
X^R_z &=& -x^{-1}\parcial{z} \,.
\end{array}
\end{equation}

The commutations relations are given by:
\begin{eqnarray}
[X^L_x,X^L_y] &=& 2X^L_y \nn \\
{[X^L_x,X^L_z]} &=& 2X^L_z  \\
{[X^L_y,X^L_z]}&=& 0\nn\,.
\end{eqnarray}

 This algebra admits a non polynomial (and non smooth) Casimir, given by:
\begin{equation}
\hat{C} = \frac{X^{L,R}_y}{X^{L,R}_z}\,.
\end{equation}

The momentum maps for the left and right action of $G$ on $T^*G$ are:
\begin{equation}
\begin{array}{rcl}
\calP^L_x &=& xp_{x} -yp_{y} -zp_{z} \\
\calP^L_y &=& xp_{y} \\
\calP^L_z &=& xp_{z}
\end{array} \qquad\qquad
\begin{array}{rcl}
\calP^R_x &=& -xp_{x} -yp_{y} -zp_{z} \\
\calP^R_y &=& -x^{-1}p_{y} \\
\calP^R_z &=& -x^{-1}p_{z} \,.
\end{array}
\end{equation}

The classical Casimir is given by
$C=\frac{\calP^{L,R}_y}{\calP^{L,R}_z} = \frac{p_y}{p_z}$.

Co-adjoint orbits are given, as usual, by the co-adjoint action of
the group on $\calG^*$, parametrized by $(p_x,p_y,p_z)$. In this
case there are orbits given by constant values of the Casimir
function on $\calG^*$, and orbits given by invariant relations
(functions $f_i$ on $\calG^*$ which are not invariant under the
co-adjoint action but such that the equations $f_i=0$ are
preserved) \cite{Levi-Civita}. The invariant relations are
$p_y=0,p_z=0$, determining the points $(p_x,0,0)$, for $p_x\in R$
as zero dimensional co-adjoint orbits.

The co-adjoint orbits given by the constant values of the Casimir
function are given by $p_y/p_z = c$, which, for convenience we
denote as  $c=\tan\theta$. These orbits are planes distributed as
a "book", i.e. all of their closures "intersect" along the line
$p_y=p_z=0$. This line, however, does not lies in these co-adjoint
orbits, since each point of the line is a co-adjoint orbit by
itself. It is convenient to parametrize these planes in the form
$(p_x,r\sin\theta,r\cos\theta)$, with $r\neq 0$, where we can
interpret $\theta$ as the angle between the plane and the
$p_y$-axis. As in the case of the 2-dim group (isomorphic to
$SB(2,R)$), these orbits are disconnected ($r>0$ and $r<0$) if we
consider the action of $G$ or each one decomposes in two connected
co-adjoint orbits if we consider $G^+$.

The Poisson structure can be inverted in each 2-dimensional
co-adjoint orbit, providing a symplectic structure of the form:
\begin{equation}
w_{\theta} = \frac{1}{2r} dp_x\wedge dr\,.
\end{equation}

Clearly, canonical coordinates are given by the Darboux map
$S(p_x,p_y,p_z)=(\frac{p_x}{2r},r)=(\pi,q)$. This allows to define
canonical coordinates in $T^*G$ by means of the pull-back by left and right
invariant momenta:
\begin{eqnarray}
X&=& (\calP^L)^*{\cdot} S^*(q) = \frac{\calP^L_y}{\sin\theta}=
           x\sqrt{p_y^2+p_z^2} \nn \\
P_X &=& (\calP^L)^*{\cdot} S^*(\pi)=\frac{\calP^L_x}{2\calP^L_y}\sin\theta =
\frac{xp_x-yp_y-zp_z}{2x\sqrt{p_y^2+p_z^2}} \nn \\
 & & \\
Y&=&(\calP^R)^*{\cdot} S'{}^*(q') =
\frac{\calP^R_2}{\sin\theta}= \frac{\sqrt{p_y^2+p_z^2}}{x} \nn  \\
P_Y &=& (\calP^R)^*{\cdot} S'{}^*(\pi')=\frac{\calP^R_x}{2\calP^R_y}\sin\theta =
\frac{x(xp_x+yp_y+zp_z)}{2\sqrt{p_y^2+p_z^2}}\nn \,,
\end{eqnarray}

\ni where we have replaced $\tan\theta$ with $p_y/p_z$, and used
the fact that it remains unchanged under the left and right
pull-backs. Inverting theses relations we obtain:
\begin{eqnarray}
\calP^L_x &=& 2XP_X \nn \\
\calP^L_y &=& \frac{P_Z}{\sqrt{1+P_Z^2}}X =\sin\theta X\\
\calP^L_z &=& \frac{1}{\sqrt{1+P_Z^2}}X= \cos\theta X \nn\,,
\end{eqnarray}

\ni where $P_Z\equiv \tan\theta$, and identical expressions for the
$\calP^R$'s in terms of $P_Y$, $Y$ and $P_Z$.

To complete the canonical transformation from $(x,y,z,p_x,p_y,p_z)$ to
$(X,Y,Z,P_X,P_Y$, $P_Z)$, we need the expression of the cyclic variable $Z$. This
is easily obtained to be:
\begin{equation}
Z=\frac{yp_z-zp_y}{1+P_Z^2}= \frac{p_z^2(yp_z-zp_y)}{p_y^2+p_z^2}\,.
\end{equation}

Let us compute the irreducible representations using the procedure
explained in Sec. \ref{Irreducible}. For each 2-dimensional
co-adjoint orbit, characterized by the Casimir $c=\tan\theta$, we
choose a point $p=(0,\sin\theta,\cos\theta)$. The hypersurface
$\Sigma^R$ associated with it is given by
$\Sigma^R=\calP^L{}^{-1}(p)=\{(x,y,z,p_x,p_y,p_z)\in T^*G\ {\rm
such\ that\ }
xp_x+yp_y+zp_z=0,\,p_y/x=\sin\theta,\,p_y/x=\cos\theta\}$. The
coordinates which parametrize $\Sigma^R$ are $X=x^2$,
$P_X=-(y\sin\theta+z\cos\theta)/x$ and
$Z=x\cos^2\theta(y\cos\theta-z\sin\theta)$.

The vertical lift of the momentum maps are given by:
\begin{eqnarray}
\hat{\calP}^L_x &=& -2iX\parcial{X}+2iP_X\parcial{P_X}\nn \\
\hat{\calP}^L_y &=& i\sin\theta \parcial{P_X} +
                     \sin\theta XI \label{SB(2,C)-operators}\\
\hat{\calP}^L_z &=& i\cos\theta\parcial{P_X} + \cos\theta XI \nn \,.
\end{eqnarray}

Since the momentum maps are polynomials up to first order in the momentum $P_X$ and the
coordinate $X$, both polarization subalgebras $\calP^X$ and $\calP^{P_X}$
are admissible, leading to equivalent representations related by the Fourier
transform. Imposing, for instance, $\calP^X$, we obtain a Hilbert space
of sections depending only of $X$, with the action of the operators given by:
\begin{eqnarray}
\hat{\calP}^L_x\Psi(X) &=& -i2X\parcial{X} \Psi(X) \nn \\
\hat{\calP}^L_y\Psi(X) &=& X\sin\theta \Psi(X) \\
\hat{\calP}^L_z\Psi(X) &=& X\cos\theta\Psi(X)\,. \nn
\end{eqnarray}

 This irreducible representation is very similar to the one obtained for
the 2-dimensional group, and as happened there, it is not unitary
with respect to the measure $dX$, and needs to be corrected with
the Radon-Nikodym derivative, which changes the first operator to
its correct expression $\hat{\calP}^L_x\Psi(X) =
-i2(X\parcial{X}+\medio) \Psi(X)$. The representations for
different values of $\tan\theta$ are, in this case, equivalent,
since there are unitary operator relating all of them. These
operators are rotations in the angle $\theta$, and are associated
with the rotations in $\calG^*$ which relate all 2-dimensional
co-adjoint orbits. These operators, as happens with the Fourier
transform, are not inner operators.

 As it happens with the 2-dimensional group, there exist another polarization
condition which does not come from any polarization subalgebra. It is
given by the operator $\parcial{P_X}-iXI$, which leads to a Hilbert subspace
preserved by the group $SB(2,C)$, where the sections have the form
$\Psi=e^{iXP_X}\Phi(X)$. The action of the operators on this sections are given
by:
\begin{eqnarray}
\hat{\calP}^L_x(e^{iXP_X}\Phi(X)) &=& e^{iXP_X}(-i2X\parcial{X})\Phi(X) \nn \\
\hat{\calP}^L_y(e^{iXP_X}\Phi(X)) &=& 0\\
\hat{\calP}^L_z(e^{iXP_X}\Phi(X)) &=& 0\,. \nn
\end{eqnarray}

This representation is irreducible and unitary with respect to the
measure $dX/X$ (or with the measure $dX$ but corrected with the
Radon-Nikodym derivative), is the same for all 2-dimensional
co-adjoint orbits, and it is not equivalent to the previous one
for the same reason as in the case of the 2-dimensional group.

\subsection{The $SL(2,R)$ group}

The $SL(2,R)$ group is the group of real $2{\times} 2$ matrices of
determinant one:
\begin{equation}
G=\left\{\left( \begin{array}{cc}
\alpha & \beta \\
\gamma & \delta
\end{array} \right),\; \hbox{such that } \alpha\delta-\beta\gamma=1\right\}\,.
\end{equation}

\noindent We will use a Gauss decomposition for $SL(2,R)$ of the form:
\begin{equation}
\left( \begin{array}{cc}
\alpha & \beta \\
\gamma & \delta
\end{array} \right) =
\left( \begin{array}{cc}
1 & 0 \\
y & 1
\end{array} \right)
\left( \begin{array}{cc}
e^z & 0\\
0 & e^{-z}
\end{array} \right)
\left( \begin{array}{cc}
1 & x \\
0 & 1
\end{array} \right)\,.
\end{equation}

\noindent The group law $g''=g'*g$ for this group in terms of the variables $x,y,z$
is given by:
\begin{eqnarray}
x'' &=& x + e^{-2z}\frac{x'}{1+x'y} \nn \\
y''&=&y'+e^{-2z'}\frac{y}{1+x'y}\\
z'' &=& z' + z + \log(1+x'y)\nn \,.
\end{eqnarray}

Left and right-invariant vector fields are given by:
\begin{equation}
\begin{array}{rcl}
X^L_x &=& \parcial{x} \\
X^L_y&=&e^{-2z}\parcial{y}-x^2\parcial{x}+x\parcial{z}\\
X^L_z&=&\parcial{z}
- 2x\parcial{x}
\end{array}\hbox{\hskip 1cm}
\begin{array}{rcl}
X^R_x &=& -e^{-2z}\parcial{x}+y^2\parcial{y}-y\parcial{z} \\
X^R_y&=&-\parcial{y}\\
X^R_z&=&-\parcial{z}+2y\parcial{y}\,.
\end{array}
\end{equation}

The commutation relations are:
\begin{eqnarray}
[X^L_x,X^L_y] &=& X^L_z \nn \\
{[X^L_x,X^L_z]} &=& -2X^L_x  \\
{[X^L_y,X^L_z]}&=& 2X^L_y \nn\,.
\end{eqnarray}

This Lie algebra admits a Casimir, which is given by the second order operator
\begin{equation}
\hat{C}_2 = (X^{L,R}_z)^2 + 2(X^{L,R}_x X^{L,R}_y+ X^{L,R}_y X^{L,R}_x)\,.
\end{equation}

We compute the left and right-invariant momentum maps
$\calP^{L,R}:T^*G\rightarrow \calG^*$ as usual, $\calP^{L,R}_i\equiv
i_{X^{L,R}_i}\Theta_0$, where $\Theta_0$ is the canonical 1-form on $T^*G$,
$\Theta_0=  p_x dx + p_y dy + p_zdz$. We obtain:
\begin{equation}
\begin{array}{rcl}
\calP^L_x &=& p_{x} \\
\calP^L_y &=& e^{-2z}p_{y} -x^2p_x+ xp_{z}\\
\calP^L_z &=& p_z -2xp_x
\end{array}\hbox{\hskip 1cm}
\begin{array}{rcl}
\calP^R_x &=& -e^{-2z}p_{x}+y^2p_y-yp_z \\
\calP^R_y &=& -p_{y}\\
\calP^R_z &=& -p_z+2yp_y\,.
\end{array}
\end{equation}

The classical Casimir is given by $C_2=(\calP^{L,R}_z)^2 +
4\calP^{L,R}_x\calP^{L,R}_y = p_z^2+4e^{-2z}p_xp_y$.

To find the canonical variables $(X,Y,Z,P_X,P_Y,P_Z)$ in $T^*G$ we
proceed as explained before, computing Darboux coordinates on
co-adjoint orbits. For the case of $SL(2,R)$, co-adjoint orbits
are characterized by the positive, null or negative values of the
classical Casimir $C_2$. Each positive value of the Casimir
corresponds to a single two-dimensional orbit (a one-sheet
hyperboloid), each negative value of the Casimir corresponds to
two two-dimensional orbit (the two-sheet hyperboloid), meanwhile
the zero value of the Casimir is associated with three orbits, the
origin (zero-dimensional) and the upper and lower sheets of the
cone.

As before, we shall restrict to the two-dimensional orbits (the
only zero-dimensional orbit, the origin in $\calG^*$, is
associated with the trivial representation, the only
one-dimensional representation of $SL(2,R)$). We shall consider
first the cases of the 1-sheet hyperboloid (with positive Casimir)
and the two cones. The 2-sheet hyperboloids will be consider
later.


Define, for $C_2\geq 0$, $R\equiv \sqrt{C_2}$, then the 2-form $\omega$ for
the 1-sheet hyperboloid and cone orbits, obtained by inverting the Poisson
brackets on the orbits, can be written as
\begin{equation}
\omega=\frac{1}{2p_x}dp_z\wedge dp_x\,,
\end{equation}

\noindent and Darboux coordinates are given by the map
$S:\calG^*\rightarrow T^*R$:
\begin{equation}
S(p_x,p_y,p_z)=(p_x,\frac{R-p_z}{2p_x})=(\pi,q)\,.
\end{equation}

It is convenient to introduce a second set of Darboux
coordinates given by the map $S':\calG^*\rightarrow T^*R$:
\begin{equation}
S'(p_x,p_y,p_z)= (\frac{R-p_z}{2p_y},-p_y)= (\pi',q')\,.
\end{equation}

These maps parametrize the 1-sheet hyperboloid when $R>0$. When
$R=0$ we have three different orbits, according to the values
$\pi=0$ (zero-dimensional orbit, the origin in $\calG^*$) and the
two cones given by $\pi<0$ and $\pi>0$.

 Once we have found Darboux coordinates on the co-adjoint orbits, we
simply define $X,P_X$ (resp. $Y,P_Y$) as the pullback by $\calP^L$ (resp.
$\calP^R$) of $S^*(q)$ and $S^*(\pi)$ (resp. of $S'{}^*(q')$ and
$S'{}^*(\pi')$), respectively:
\begin{eqnarray}
X&=& (\calP^L)^*{\cdot} S^*(q) = \frac{R-\calP^L_z}{2\calP^L_x}=
           \frac{R-p_z+2xp_x}{2p_x}\nn \\
P_X &=& (\calP^L)^*{\cdot} S^*(\pi)=\calP^L_x = p_x \nn \\
 & & \\
Y&=&(\calP^R)^*{\cdot} S'{}^*(q') =
\frac{R-\calP^R_z}{2\calP^R_y}= -\frac{R+p_z-2yp_y}{2p_y} \nn  \\
P_Y &=& (\calP^R)^*{\cdot} S'{}^*(\pi')=-\calP^R_y = p_y \nn \,,
\end{eqnarray}

\ni where $R$, since it is a function of the Casimir $C_2$, it remains
unchanged to the value $R=\sqrt{p_z^2 + 4e^{-2z}p_xp_y}$. Inverting these
relations we obtain the expressions of $\calP^{L,R}_i$ in terms of $X$ and
$P_X$ (resp. $Y$ and $P_Y$):
\begin{equation}
\begin{array}{rcl}
\calP^L_x &=& P_X  \\
\calP^L_y &=& XP_Z-X^2P_X \\
\calP^L_z &=& P_Z -2XP_X \\
\end{array}\hbox{\hskip 1cm}
\begin{array}{rcl}
\calP^R_x &=& Y^2P_Y-YP_Z \\
\calP^R_y &=& - P_Y  \\
\calP^R_z &=& 2YP_Y-P_Z \,,
\end{array}
\end{equation}

\ni  where $P_Z$
has been introduced, being the function of the Casimir $P_Z=R=\sqrt{C_2}$.
This expressions
are clearly linear in the momenta, so they can be properly ``quantized".
The only thing we have
to compute is the cyclic coordinate $Z$ canonically conjugated to $P_Z$.
As in the previous
example, using the fact that the canonical 1-form $\Theta_0$ in $T^*G$ can
be written as
$P_XdX+P_YdY+P_ZdZ$, we derive the expression for $Z$:
\begin{equation}
Z=\log \frac{p_x}{p_y}+ \tanh^{-1}\frac{p_z}{P_Z}\,.
\end{equation}


Let us compute the irreducible representations associated with the
1-sheet hyperboloid and cone co-adjoint orbits. In each orbit, we
choose $p=(R/2,R/2,0)$, and the hypersurface $\Sigma^R$ associated
with it is given by
$\Sigma=\calP^L{}^{-1}(p)=\{(x,y,z,p_x,p_y,p_z)\in T^*G\ {\rm
such\ that\ }
e^{-2z}p_x-y^2p_y-yp_z=R/2,\,p_y=R/2,\,p_z-2yp_y=0\}$. This
hypersurface is parameterized by the coordinates
$X=x+\frac{e^{-2z}}{1+y}$, $P_X=\frac{R}{2}e^{2z}(1-y^2)$ and $Z=
2z + \log(1-y^2) + \tanh^{-1}y$.

The vertical lift of the left invariant momentum maps are given by:
\begin{eqnarray}
\hat{\calP}^L_x &=& -i\parcial{X}\nn \\
\hat{\calP}^L_y &=& i X^2 \parcial{X} - 2iXP_X\parcial{P_X} +
                     X P_Z I \label{SL(2,R)-operators}\\
\hat{\calP}^L_z &=& 2iX\parcial{X} -2iP_X\parcial{P_X}+ P_ZI \nn \,.
\end{eqnarray}

Since the left-invariant momentum maps are at most of first order in the
momentum $P_X$, the polarization subalgebra $\calP^X$ is admissible.
However, $\calP^{P_X}$ is not admissible since they are of second order in
$X$. The Hilbert space $\calH^X$ is made of  sections depending only on
$X$, and the operators (\ref{SL(2,R)-operators}) reduce to:
\begin{eqnarray}
\hat{\calP}^L_x\Psi(X) &=& -i\parcial{X} \Psi(X) \nn \\
\hat{\calP}^L_y\Psi(X) &=& (iX^2\parcial{X}+XP_Z) \Psi(X)  \\
\hat{\calP}^L_z\Psi(X) &=& (2iX\parcial{X} + P_Z)\Psi(X)\,. \nn
\end{eqnarray}

This representation is irreducible but not unitary with respect to the
scalar product given by the measure $dX$. It requires the addition
of the Radon-Nikodym derivative, which transforms the operators into:
\begin{eqnarray}
\hat{\calP}^L_x\Psi(X) &=& -i\parcial{X} \Psi(X) \nn \\
\hat{\calP}^L_y\Psi(X) &=& [iX^2\parcial{X}+X(i+P_Z)] \Psi(X)  \\
\hat{\calP}^L_z\Psi(X) &=& [2iX\parcial{X} + (i+P_Z)]\Psi(X)\,, \nn
\end{eqnarray}

\ni providing a unitary representation of $SL(2,R)$. The value of the Casimir
operator in each representation is $\hat{C}_2=1+P_Z^2$.

 There are no other polarization conditions, therefore the only irreducible
unitary representations for each value of $P_Z^2>0$ are the ones
presented here, which correspond to the continuous series of
representations of $SL(2,R)$.

\section{Conclusions}

We have shown that the irreducible representations of Lie groups
can be constructed relating the classical configuration space (G
group) and classical phase space ${T^*G}$, in which the orbits of
the group action on the points in the spaces are defined, to the
Lie group. The quantization procedure, which uses canonical
quantization replacing the classical momenta in the phase space by
the standard momentum operators, provides the construction of the
generators of the Lie group representation. In quantum mechanics
and quantum optics, the presented geometrical picture of the Lie
group representations clarifies the group-theoretical meaning of
 the quadrature  components and their statistical properties associated with
 different basis vectors in the representation on Hilbert space.
 One should point out that in the mathematical context the Lie group
 representation theory has been constructed long time ago, but to apply
 this formalism in quantum mechanics and quantum optics, one needs a tutorial
 presentation of the Lie groups and their irreducible representations.
 The geometrical picture developed in this work, which treats the T*G as the
 phase space associated to the G group and connects the irreducible
 representations with group ``trajectories'' in this phase space,
provides a conventional tool to access the rigorous mathematics to
 physical intuition. The considered examples of Lie groups
of low dimensions show the properties of the group representations
in visible physical images like positions and momenta and their
change in the process of evolution. The evolution itself is
treated as a simple trajectory in the phase space of quadratures
under the action of the evolution operator obtained in terms of
very simple Hamiltonian which is linear in the position operator.

\end{document}